\documentstyle[twocolumn,aps,psfig,floats]{revtex}

\begin{document}

\draft

\title{Large sample-to-sample fluctuations 
of the nonequilibrium critical current through mesoscopic Josephson junctions}
\author{P. Samuelsson$^{a}$ and H. Schomerus$^{b}$}

\address{
$^a$ Department of Microelectronics and Nanoscience,
Chalmers University of Technology
and
G\"{o}teborg University, \\S-41296 G\"{o}teborg, Sweden\\
$^b$ Instituut-Lorentz, Universiteit Leiden, P.O. Box 9506, 2300 RA Leiden,
The Netherlands}
\date{\today}
\twocolumn[
\widetext
\begin{@twocolumnfalse}

\maketitle

\begin{abstract}
        We present a theory for the nonequilibrium current in a mesoscopic Josephson 
        junction which is coupled to a normal electron reservoir,
        and apply it to a chaotic junction.
        Large sample-to-sample fluctuations of the critical current
        $I_{\rm c}$ are found, with $\mbox{rms}~I_{\rm c}
        \simeq \sqrt{N}e\Delta/\hbar$, 
        when the voltage difference $eV$ between the 
        electron reservoir and the junction exceeds the superconducting gap
        $\Delta$ and the number of modes $N$ connecting the junction to
        the superconducting electrodes is large.
\end{abstract}

\pacs{PACS: 74.50.+r, 74.20.Fg, 74.80.Fp}
\end{@twocolumnfalse}
]

\narrowtext
Over the last years there has been an increased interest in the
nonequilibrium Josephson current in mesoscopic multiterminal
superconductor-normal metal-superconductor (SNS)
junctions. Nonequilibrium in the junction is created by
quasiparticle injection from one or several normal electron reservoirs,
connected to the normal part of the SNS junction. By controlling the
voltage applied between the normal reservoirs and the SNS junction, it
has been shown in recent experiments that the Josephson current can be
suppressed \cite{morpurgo98,schapers98}, reversed \cite{baselmans99}
and in the case with injection from a superconducting reservoir, even
enhanced \cite{kutchinsky99}.

The microscopic mechanism for these effects, nonequilibrium population
of the current-carrying Andreev levels, was discussed by van Wees et
al.\ \cite{vanwees91} already in 1991. Thereafter, the nonequilibrium
Josephson current in various multiterminal geometries has been studied
in both diffusive \cite{volkov95,yip98,wilhelm98} and
quantum ballistic \cite{chang97,samuelsson97} junctions. In
Ref.\ \cite{samuelsson97} it was pointed out that the nonequilibrium
Josephson current in ballistic SNS junctions can not be described only
in terms of the nonequilibrium population of Andreev levels: There is
also a quantum-interference addition to the Josephson current,  
which results from the difference between
the scattering-state wavefunctions for injected electrons and holes.

\begin{figure}
\centerline{\psfig{figure=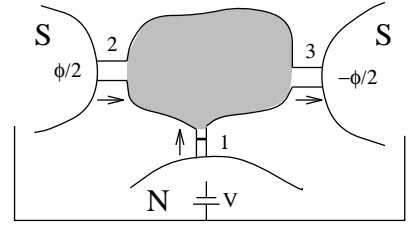,width=0.6\linewidth}}
\medskip
\caption{
        Three-terminal SNS junction, consisting
        of a mesoscopic
  scatterer (grey shaded) connected to two superconducting
  reservoirs via contacts $2$ and $3$ and a normal
  reservoir via contact $1$.
        The black bar in contact $1$ indicates a tunnel barrier,
  the arrows the direction of positive current flow.}
\label{junction}
\end{figure}

In this paper we develop a general theory of the nonequilibrium
Josephson current
in three-terminal SNS junctions (see Fig.\ \ref{junction}),
in terms of the scattering matrix  
for electrons and holes injected 
from the normal reservoir. The theory is then applied to a chaotic 
junction, in the limit of weak coupling to the normal reservoir
and at zero temperature.
We find 
that the quantum-interference
contribution gives rise to
sample-to-sample fluctuations of the critical current $I_{\rm c}$
which are much larger than the
equilibrium fluctuations \cite{beenakker91,takayanagi95}:
For a large voltage $V$ (with $eV\gtrsim\Delta$, the superconducting gap),
\begin{equation}
\mbox{rms}~I_{\rm c}\equiv \sqrt{\langle I_{\rm c}^2\rangle-\langle I_{\rm c}\rangle^2}
\simeq \langle I_{\rm c}\rangle
\simeq \sqrt{N}\frac{e\Delta}{\hbar},
\label{ic}
\end{equation}
hence the fluctuations are
of the order of the ensemble-averaged critical current itself.
(Here 
$N$ is the number of modes connecting the junction to
each of the superconducting electrodes.)
In this regime the current results from the 
quantum-interference
contribution alone.
For $eV\lesssim \Delta$ the critical current
is of order $N\frac{e\Delta}{\hbar}$, with
fluctuations of order $\frac{e\Delta}{\hbar}$.

A model of the junction is presented in Fig.\ \ref{junction}. A
mesoscopic scatterer is connected to two superconducting
leads via ballistic contacts, each supporting $N$ transverse
modes. The phase difference between the superconductors is
$\phi$.
The scatterer is also connected to a normal reservoir via a contact
with $M$ modes, containing a tunnel barrier with
transparency $\Gamma$.
A voltage $V$ is applied
between the SNS junction and the normal reservoir.
We assume that the resistance of
the injection contact is the dominating resistance of the junction,
such that
the potential drops completely over the injection point. In order to preserve
nonequilibrium, the strength of the tunnel barrier $\Gamma$ is however
limited by the requirement that the dwell time of the injected
quasiparticles $\propto 1/\Gamma$ must be smaller than the inelastic
scattering time in the junction.

Under these conditions, the distribution of the quasiparticles in the
junction is determined by the distributions
$n^{\rm e(h)}=n_{\rm F}(E\mp eV)$
of electrons (holes) in the reservoir, where
$n_{\rm F}=[1+\exp(E/kT)]^{-1}$.
The current in contact $j=1$, $2$, $3$ can then be written as
\begin{equation}
I_j=\int_{-\infty}^{\infty}dE(i^{\rm e}_j n^{\rm e}+i^{\rm h}_j n^{\rm h}+i^{\rm
s}_j n_{\rm F})
,
\label{ijdef}
\end{equation}
with $i_j^{\rm e(h)}$ the current density of the scattering states
resulting from injected electron (hole) quasiparticles from the normal
reservoir and $i^{\rm s}_j$ the total current density for quasiparticles
injected from the superconductors ($i^{\rm s}=0$ for subgap energies $|E|<\Delta$).

The current $I= (I_2+I_3)/2$
flowing between the superconductors
can be rewritten by 
using the current conservation for each energy
$i_1^{\rm e,h}+i_2^{\rm e,h}=i_3^{\rm e,h}$ and the fact that no current is
flowing in the injection lead in equilibrium, $i_1^{\rm e}+i_1^{\rm h}=0$.
It takes then the form $I=I^{\rm neq}+I^{\rm eq}$,
where the  equilibrium current at $eV=0$ is given by
$I^{\rm eq}=\int dE[
i^+ +(i^{\rm s}_2+i^{\rm s}_3)/2]
n_{\rm F}$, and
\begin{equation}
I^{\rm neq}=\int\limits_{-\infty}^{\infty}dE\,
\left[\frac{i^+}{2}(n^{\rm e}+n^{\rm h}-2n_{\rm F})
+\frac{i^-}{2}(n^{\rm e}-n^{\rm h})\right]. 
\label{ineq}
\end{equation}
Here the current densities $i^+=i_2^{\rm e}+i_2^{\rm h}=i_3^{\rm e}+i_3^{\rm h}$ and
$i^-=(i_2^{\rm e}-i_2^{\rm h}+i_3^{\rm e}-i_3^{\rm h})/2$
are the sum and the difference of the current densities of
the scattering states for injected electrons and holes.
The contribution $\propto i^+$ to $I^{\rm neq}$ results from the nonequilibrium population of
the Andreev levels,
while the current $\propto i^-$
accounts for the quantum-interference contribution as
well as for an asymmetric splitting  of the injected current
$I_1=\int dE(i_1^{\rm e}-i_1^{\rm h})(n^{\rm e}-n^{\rm h})/2$.

We will now
express the current densities in terms of the scattering matrix $S$ of
injected quasiparticles from the reservoir \cite{beenakker91}.
The current densities are calculated
most conveniently in the contacts $j=1$, $2$, $3$, where the wavefunctions are
plane-wave solutions to the Bogoliubov-de Gennes equation.
A wave incident on the scatterer from leads 2 and 3 is described by the $4N$
vector of wavefunction coefficients
$c_{\rm in}=(c_2^{{\rm e},+},c_3^{{\rm e},-},c_2^{{\rm h},-},c_3^{{\rm h},+})$.
The superscript
$+(-)$  denotes a positive
(negative) sign of the wave vector. Correspondingly the outgoing wave
is given by
$c_{\rm out}=(c_2^{{\rm e},-},c_3^{{\rm e},+},c_2^{{\rm h},+},c_3^{{\rm h},-})$. At
the NS interfaces, Andreev reflection is described by the
scattering matrix
\begin{equation}
S_{\rm A}=\alpha \left( \begin{array}{cc} 0 &
r_{\rm A} \\ r_{\rm A}^* & 0 \end{array} \right),
r_{\rm A}=\left( \begin{array}{cc} e^{i\phi/2} & 0 \\ 0 &
e^{-i\phi/2} \end{array} \right) 
,
\end{equation}
such that $c_{\rm in}=S_{\rm A} c_{\rm out}$, with
$\alpha=\exp[-i\arccos(E/\Delta)]$.
The wavefunctions
in the three contacts are then matched with help of the $(2N+M)\times(2N+M)$
scattering matrix $S'$ of the normal region (including the tunnel barrier),
\begin{equation}
S'=\left( \begin{array}{ccc} r_{11} & t_{12} & t_{13} \\ t_{21}
& r_{22} & t_{23} \\ t_{31} & t_{32} & r_{33} \end{array} \right)
.
\label{nscatmat}
\end{equation}
We introduce a non-unitary
matrix $S_{\rm N}$, describing only the scattering between the
contacts $j=2$ and $3$, 
\begin{equation}
S_{\rm N}=\left( \begin{array}{cc} S_0(E) & 0 \\ 0 & S_0^*(-E) \end{array} \right),\quad
S_0=\left( \begin{array}{cc} r_{22} & t_{23} \\ t_{32} & r_{33} \end{array} \right)
,
\label{sndef} 
\end{equation}
such that $c_{\rm out}=S_{\rm N} c_{\rm in}$,
and matrices which involve also contact $1$,
\begin{eqnarray*}
{\cal T}&=&\left( \begin{array}{cccc}
t_{12}(E) & t_{13}(E) & 0 &  0 \\ 0 & 0 & t^*_{12}(-E) & t^*_{13}(-E) \end{array}
\right), \\
{\cal T}'&=& \left( \begin{array}{cc} t_{21}(E) & 0 \\ t_{31}(E) & 0 \\ 0 &  t^*_{21}(-E)
\\ 0  & t^*_{31}(-E) \end{array} \right),~ {\cal R}=\left( \begin{array}{cc} r_{11}(E) &
0 \\ 0  & r^*_{11}(-E) \end{array} \right). \nonumber
\end{eqnarray*}
>From these
ingredients, the
coefficients $c$ can be calculated and the current densities
in Eq.\ (\ref{ijdef}) are obtained from the quantum
mechanical expression for current. The current densities
$i^+$ and $i^-$ follow after some matrix algebra,
\begin{eqnarray}
i^{+}(E)&=&\frac{e}{h}\mbox{tr}\left({\cal T}'^{\dagger}
(1-S_{\rm A}^{\dagger}S_{\rm N}^{\dagger})^{-1} \sigma_z'
(1-S_{\rm N}S_{\rm A})^{-1}{\cal T}'\right)
,
\nonumber
\\
i^{-}(E)&=&\frac{e}{h}\mbox{tr}\left({\cal T}'^{\dagger}(1-S_{\rm A}^{\dagger}
S_{\rm N}^{\dagger})^{-1} \sigma_z'(1-S_{\rm N}S_{\rm A})^{-1}{\cal T}'\sigma_z\right)
,
\nonumber
\\
\label{iminuseq}
\end{eqnarray}
where  $\sigma_z={\rm diag}~(1,1,-1,-1)$,
$\sigma_z'={\rm diag}~(-1,1,1,-1)$.
The scattering matrix $S$ for injected
quasiparticles from the normal reservoir can be written as
\begin{equation}
S=\left( \begin{array}{cc} r_{\rm ee} & r_{\rm he} \\ r_{\rm eh}  & r_{\rm hh}
\end{array} \right)={\cal R}+{\cal T}(S_{\rm A}^{\dagger}-S_{\rm N})^{-1}{\cal T}'.
\label{smat}
\end{equation}
After some further matrix algebra we obtain from Eqs.\ (\ref{iminuseq}) and
(\ref{smat}) the expressions (for subgap energies $|E|<\Delta$)
\begin{equation}
i^{+}(E)=\frac{2e}{ih}\mbox{tr}\left(S^{\dagger}\frac{d}{d\phi}S\right),~~
i^{-}(E)=\frac{2e}{ih}\mbox{tr}\left(S^{\dagger}\frac{d}{d\phi}S \tau_z\right)
,
\label{ipm}
\end{equation}
with $\tau_z={\rm diag}~(1,-1)$.
(The expression for $i^+$ is well known \cite{akkermans91,brouwer96}.)
Eqs.\ (\ref{ineq}) and (\ref{ipm}) are our general results for
the nonequilibrium Josephson current.

In general, the current flowing between the superconductors contains also
the part of the injected current which is
asymmetrically split between contacts $2$ and $3$ \cite{volkov96}.
This is not the case when
the SNS junction is weakly coupled to the reservoir ($\Gamma \ll 1$),
because the injected current is then negligible.
In this limit the matrix $S_{\rm N}=S_{{\rm N}0}+\Gamma \delta S_{\rm N}$ can be
expanded to first order in $\Gamma$, where $S_{{\rm N}0}$ is unitary.
The two current densities $i^+$ and $i^-$
have the same discrete spectrum of Andreev levels,
given by the solutions $E_n$ of
$\mbox{det}(1-S_{\rm A}S_{{\rm N}0})=0$, but different spectral weights.
The current density $i^+$ reduces to the
well-known expression for the closed junction,
\begin{equation}
i^{+}(E)=\sum_n I_n^+ \delta(E-E_n),
\qquad
I_n^+=\frac{2e}{\hbar}\frac{dE_n}{d\phi}
.
\label{bstcurr}
\end{equation}
The current density $i^-$
can be found from first-order perturbation theory in the
tunnel-barrier transparency $\Gamma$,
\begin{mathletters}
\begin{eqnarray}
&&i^{-}(E)=\sum_n I_n^-\delta(E-E_n),\qquad I^-_n=R_nI^+_n,
\\
&&R_n=\frac{\mbox{Re}(U^\dagger\sigma_z \delta S_{\rm N} S_{\rm A} U)_{nn}}{
\mbox{Re}(U^\dagger \delta S_{\rm N} S_{\rm A} U)_{nn}}
,
\end{eqnarray}
\label{bstcurr2}%
\end{mathletters}
where
the unitary matrix $U$ diagonalizes the
unitary matrix product $S_{\rm A} S_{{\rm N}0}=U \mbox{diag}(\lambda)U^{\dagger}$.
One can show with help of the corresponding eigenvalue equation that
the ratios $|R_n|\le 1$.
It should be pointed out that the matrix $\delta S_{\rm N}$ can {\em not} be
expressed in terms of the closed junction scattering matrix $S_{{\rm N}0}$,
i.\,e.\ the current density $i^-$ depends manifestly on 
the properties of the contact between the normal
reservoir and the SNS junction.

\begin{figure}
\centerline{\psfig{figure=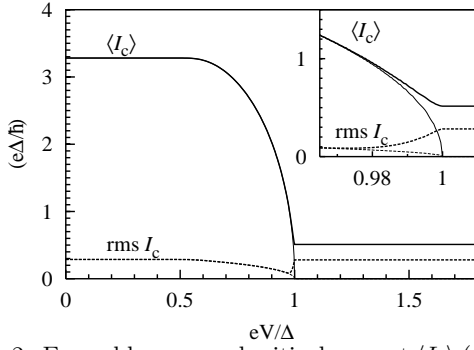,width=0.8\linewidth}}
\caption{Ensemble-averaged critical current $\langle I_{\rm c} \rangle$
(solid thick line) and the fluctuations $\mbox{rms}~I_{\rm c}$
(dashed thick line)
as a function of voltage $V$ between the normal reservoir
and the junction. The thin lines are the result with $I^-=0$
in Eq.\ (\protect\ref{ishort}).
The junction has $N=10$ modes to each of the 
superconducting electrodes and $M=1$ mode to the normal reservoir.
Inset: the voltage range $0.965\Delta<eV<1.01\Delta$.
($10^3$ random matrices $S'$ have been generated).}
\label{icrms}
\end{figure}

In order to investigate the mesoscopic fluctuations of the 
nonequilibrium current in more detail
we now apply our theory to a chaotic SNS junction,
in the limit of weak coupling to the normal reservoir
\cite{brouwer96}. 
Here we only consider the simplest case, in which
the dwell time in the normal scatterer (with the superconducting leads
replaced by normal ones)
$t_{\rm dwell}<\hbar/\Delta$. (Our main
conclusions should apply also for the opposite case.)
For such a junction we can neglect the energy dependence of $S'$,
which is then distributed with the so-called Poisson
kernel $P(S')\propto|\mbox{det}(1-\langle S'^\dagger \rangle
S')|^{-(2N+M+1)}$,
where $\langle S'\rangle$ is the
ensemble-averaged scattering matrix \cite{brouwer95}.
(The magnetic field $B=0$, which gives a 
symmetric scattering matrix $S'=S'^T$.)
Furthermore, the current density for energies outside the
gap vanishes \cite{beenakker91}. Using the energy symmetries
$i^+(E)=-i^+(-E)$ and $i^-(E)=i^-(-E)$,
the total current at zero temperature,
\begin{equation}
I=-\sum_{E_n>eV} I^+_n+\sum_{E_n<eV} I^-_n \equiv I^++I^-
,
\label{ishort}
\end{equation}
can be
written as a sum over the currents $I_n^+$ and $I_n^-$ carried by the
individual Andreev levels with positive energies $E_n$.
Eq.\ (\ref{ishort}) provides a simple picture where in equilibrium
all Andreev levels carry the currents $I^+_n$. Increasing the voltage,
the Andreev levels one by one switch from $I_n^+$ to $I_n^-$
when the voltage is passing through $eV=E_n$.
At $eV \geq \Delta$,
all levels carry the current $I_n^-$.

In terms of the transmission
eigenvalues $T_n$ of the matrix $S_0$, the Andreev bound-state energies are given by
$E_n=\Delta(1-T_n\sin^2\phi/2)^{1/2}$,
hence the relation \cite{beenakker91}
\begin{equation}
I_n^+=-(e\Delta/2 \hbar)T_n\sin\phi(1-T_n\sin^2\phi/2)^{-1/2}.
\end{equation}
The statistical properties of 
the equilibrium current $I^{\rm eq}=-\sum_n I_n^+$ are known \cite{beenakker91},
with $\langle I^{\rm eq}
\rangle \simeq Ne\Delta/\hbar$ and $\mbox{rms}~I^{\rm eq} \simeq
e\Delta/\hbar$.

For $eV\ge \Delta$ the current is $I=\sum_n I_n^-=\sum_nR_nI_n^+$.
The statistics of the ratios $R_n$ follows from the construction of all
perturbations $\delta S_{\rm N}$ which are compatible with a given
$S_{{\rm N}0}$ (i.e.,
both matrices follow from the same scattering matrix of 
the open scatterer \cite{brouwer95}).
For $M=1$ such an analysis results in 
\begin{equation}
R_n=(1-T_n)^{1/2}[(\sin^2\phi/2)^{-1}-T_n]^{-1/2}
\sin{\beta_n}
,
\label{eq:rbeta}
\end{equation}
where the angles $\{\beta_n\}$ (parameterizing the coupling to the reservoir)
are independent random numbers, uniformly distributed in the interval $[0,2\pi)$.
As a consequence for fixed phase difference $\phi$ the average current
$\langle I \rangle=0$, and the fluctuations $\mbox{rms}~I\simeq
\sqrt{N}e\Delta/\hbar$ because $I^-$ is a sum of $N$ independently
fluctuating numbers $I_n^-$.
The precise value of the 
fluctuations can be calculated upon replacing the sum in 
$\langle I^2\rangle=\langle\sum_n (R_n {I_n^+})^2\rangle$
(valid due to the independence of the $\beta_n$) by an integral over the 
transmission eigenvalues, with density $\rho(T)=N\pi^{-1}[T(1-T)]^{-1/2}$. This results in
\begin{eqnarray}
&&\mbox{rms}~I=\frac{\sqrt{N}e\Delta}{2\hbar|\tan\phi/2|}
                                 \sqrt{ \sin^2\frac{\phi}{2}+\frac{8\sin^2\phi/4}{\cos\phi/2}-3 \frac{\sin^2\phi/2}{\cos\phi/2} }
\nonumber\\
&&\qquad\qquad\qquad\mbox{for }eV\ge \Delta
,
\label{eq:irms}
\end{eqnarray}
which is parametrically larger than the equilibrium fluctuations when $N\gg 1$.

Another physical quantity of interest is 
the critical current $I_{\rm c}$, the largest possible
current for a given realization. Because of $I(\phi)=-I(-\phi)$
it sometimes makes sense to restrict the
phase to $0<\phi<\pi$ and to consider the current which is 
largest in modulus;
$I_{\rm c}$ can then be positive or
negative \cite{baselmans99}. (With this definition,
the average critical current vanishes 
for $eV>\Delta$.)
In the following, however, we maximize over  $-\pi<\phi<\pi$, hence
$I_{\rm c}$ is always positive, as it is
obtained from the $I$/$V$ characteristic in experiments.
The ensemble-averaged
critical current and its fluctuations
(obtained from a numerical simulation of the random-matrix ensemble with
$N=10$ and $M=1$)
are shown in
Fig.\ \ref{icrms}, as a function of applied voltage $eV$. 
The result is compared to the contribution of $I^+$
in Eq.\ (\ref{ishort})
alone, which
only takes the nonequilibrium population of the Andreev levels into account.

For $0\le eV\lesssim 0.54\Delta$ the critical current is equal to its equilibrium value,
because at the non-fluctuating
critical phase \cite{beenakker91} $\phi_{\rm c}\simeq 2$
all bound-state energies $E_n>eV$
(in general the energies  lie in the interval $[\Delta\cos\phi/2,\Delta]$).
In the range $0.54\Delta\le eV\lesssim 0.98\Delta$ the critical
phase is determined
by the condition $\cos\phi_{\rm c}/2=eV/\Delta$
that the first Andreev bound state drops below $eV$, with only small
fluctuations due to
the high density of transmission eigenvalues $T_n\approx 1$.
Hence the critical current is $I_{\rm c}=I^{\rm eq}(\phi_{\rm c})$.
In this regime
the quantum-interference contribution $I^-$ in Eq.\ (\ref{ishort})
does not play any role
because $\langle I^+\rangle \gg \mbox{rms}~I^-$.
For a voltage $eV \approx 0.98\Delta$ very close to the gap,
$I^+$ and $I^-$ are both of order
$\sqrt{N} e\Delta/\hbar$, and the critical current
starts to deviate from what one would expect from a pure
nonequilibrium population of the Andreev levels.
(For increasing $N$ the cross-over voltage $eV \to \Delta$.)
In parallel
the fluctuations of the critical current increase.
The critical current remains constant for $eV\ge \Delta$, where it is given solely by $I^-$.

\begin{figure}
\centerline{\psfig{figure=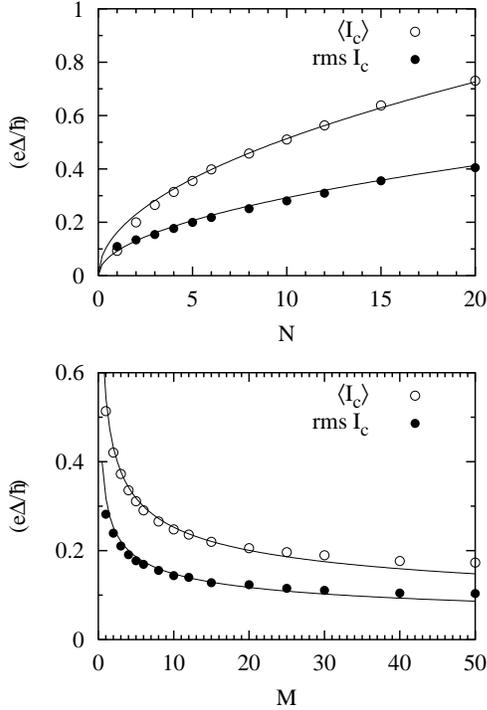,width=0.8\linewidth}}
\caption{Ensemble-averaged
critical current $\langle I_{\rm c} \rangle$
(open circles) and the fluctuations $\mbox{rms}~I_{\rm c}$ (full circles) as a
function of
the number of junction modes $N$
for a single injection mode $M=1$ (upper panel)
and
the number of injection modes $M$ for $N=10$ junction modes
(lower panel). The curves are $\propto N^{1/2}$ (upper panel)
and $\propto M^{-1/3}$ (lower panel).}
\label{rmsn}
\end{figure}

The critical current for  $eV \geq \Delta$
and its fluctuations as a function of junction modes $N$ are shown in
the upper panel of Fig.\ \ref{rmsn}.
The mean critical current is
$\langle I_{\rm c} \rangle \simeq 0.16 \sqrt{N}
e\Delta/\hbar$.
The fluctuations are of
the same order, $\mbox{rms}~I_{\rm c} \simeq 0.1 \sqrt{N} e\Delta/\hbar$,
which is by a factor
of about $\sqrt{N}/3$ larger than the equilibrium fluctuations.
Hence the $N$ dependence in Eq.\ (\ref{eq:irms}) carries over to
the average critical current and its fluctuations.

Finally let us consider the dependence of the critical current on the
number of injection modes $M$. 
This number is significant because
the current $I^-$ depends manifestly
on the coupling of the reservoir to the junction
[see Eq.\ (\ref{bstcurr2})], in
contrast to the current $I^+$ which only depends on properties of
the decoupled junction. The lower panel of Fig.\ \ref{rmsn} shows
that the critical current and its fluctuations at $eV \geq\Delta$
are suppressed when $M$
is increased. 
The
functional dependence is approximately $\propto M^{-1/3}$.
The curves flatten out
when $M$ becomes larger than the
total number $2N$ of modes connected to the superconductors. Thus, for an
experimental observation of the large fluctuations predicted above, an
injection contact with few modes is favorable.

In conclusion, we have studied the nonequilibrium Josephson current in
a mesoscopic SNS junction connected to a normal electron reservoir. It
is found that the current can be expressed in terms of the scattering
matrix for the quasiparticles injected from the normal reservoir,
Eqs.\ (\ref{ineq}) and (\ref{ipm}). As an application
we considered the nonequilibrium
current in a chaotic Josephson junction 
at zero
temperature, weakly coupled to the normal reservoir. It
is found that the fluctuations of the critical current for a
voltage $eV\geq \Delta$ are of order $\mbox{rms}~I_{\rm c} \simeq \sqrt{N}
\Delta e/\hbar$, which is of the same order as the mean critical current itself, and
much larger than the equilibrium fluctuations (of
order $\Delta e/\hbar$).

We acknowledge discussions with C. W. J. Beenakker, V. S.
Shumeiko, and G. Wendin. This work has been supported by TFR, NUTEK, NEDO, and
NWO/FOM.


\vspace*{-.5cm}
\begin{thebibliography}{99}
\vspace*{-1.6cm}

\bibitem{morpurgo98}
A. F. Morpurgo, T. M. Klapwijk, and B. J. van Wees, Appl. Phys. Lett. {\bf 72},
966 (1998).
\bibitem{schapers98}
Th. Sch{\"a}pers {\em et al.}, %
Appl. Phys. Lett. {\bf 73}, 2348 (1998).
\bibitem{baselmans99}
J. J. A. Baselmans, A. F. Morpurgo, B. J. van Wees, and T. M. Klapwijk,
Nature {\bf 397}, 43 (1999).
\bibitem{kutchinsky99} 
J. Kutchinsky {\em et al.},
Phys. Rev. Lett. \textbf{83},
4856 (1999).
\bibitem{vanwees91}
B. J. van Wees, K.-M. H. Lenssen, and C. J. P. M. Harmans,
Phys. Rev. B {\bf 44}, 470 (1991).
\bibitem{volkov95}
A. F. Volkov, Phys. Rev. Lett. \textbf{74}, 4730 (1995).
\bibitem{yip98}
S. K. Yip, Phys. Rev. B \textbf{58}, 5803 (1998).
\bibitem{wilhelm98}
F. K. Wilhelm, G. Sch\"on, and A. Zaikin, Phys. Rev. Lett. \textbf{81}, 1682 (1998).
\bibitem{chang97}
L. F. Chang and P. F. Bagwell, Phys. Rev. B {\bf 55},
12678 (1997).
\bibitem{samuelsson97}
P. Samuelsson, V. S. Shumeiko, and G. Wendin, Phys. Rev. B {\bf 56}, R5763
(1997).
\bibitem{beenakker91}
C. W. J. Beenakker, Phys. Rev. Lett. \textbf{67}, 3836 (1991);
C. W. J. Beenakker, Phys. Rev. B \textbf{74}, 15763 (1993).
\bibitem{takayanagi95}
H. Takayanagi, J. Bindslev Hansen, and J. Nitta, Phys. Rev. Lett. {\bf
74}, 166 (1995).
\bibitem{akkermans91}
E. Akkermans, A. Auerbach, J. E. Avron, and B. Shapiro, Phys. Rev. Lett. {\bf
66}, 76 (1991).
\bibitem{brouwer96}
P. W. Brouwer and C. W. J. Beenakker, Chaos, Solitons and Fractals
{\bf 8}, 1249 (1997). 
\bibitem{volkov96}
A. F. Volkov and V. V. Pavlovski{\u\i}, 
Pis'ma Zh. Eksp. Teor. Fiz. {\bf 64}, 624 (1996)
[JETP Lett. {\bf 64}, 670 (1996)];
A. F. Volkov and H. Takayanagi, Phys. Rev. B {\bf 56}, 11184 (1997).
\bibitem{brouwer95}
P. W. Brouwer, Phys. Rev. B {\bf 51}, 16878 (1995).
\end{thebibliography}
\end{document}